\documentclass[iop]{emulateapj}

\usepackage{amssymb, amsmath, amsfonts}

\usepackage{graphicx, shadow}

\usepackage{rotating}
\usepackage{wasysym}
\usepackage{verbatim}
\usepackage{natbib}

\newcommand{\ngc}{$N_{GC}$}
\newcommand{\mbh}{$M_{BH}$}

\shorttitle{Globular Cluster -- Black Hole Relation}
\shortauthors{G. Snyder et al.}

\begin{document}

\title{Relation Between Globular Clusters and Supermassive Black Holes in Ellipticals as a Manifestation of the Black Hole Fundamental Plane}
\shorttitle{Globular Cluster -- Black Hole Relation}
\shortauthors{}


\author{Gregory F. Snyder\altaffilmark{1}, Philip F. Hopkins\altaffilmark{2,3}, and Lars Hernquist\altaffilmark{1}}
\altaffiltext{1}{Harvard-Smithsonian Center for Astrophysics, 60 Gardin ST, Cambridge, MA 02138}
\altaffiltext{2}{University of California at Berkeley, Department of Astronomy, 553 Campbell Hall, Berkeley, CA 94720}
\altaffiltext{3}{Miller Fellow}

\journalinfo{Accepted to The Astrophysical Journal Letters}
\slugcomment{Submitted 2010 November 22; Accepted 2011 January 6}

\keywords{black hole physics---galaxies: evolution---galaxies: formation---galaxies: elliptical and lenticular, cD}


\begin{abstract}
We analyze the relation between the mass of the central supermassive black hole (\mbh) and the number of globular clusters (\ngc) in elliptical galaxies and bulges as a ramification of the black hole fundamental plane, the theoretically predicted and observed multi-variable correlation between \mbh\ and bulge binding energy.  Although the tightness of the \mbh--\ngc\ correlation suggests an unlikely causal link between supermassive black holes and globular clusters, such a correspondence can exhibit small scatter even if the physical relationship is indirect.  We show that the relatively small scatter of the \mbh--\ngc\ relation owes to the mutual residual correlation of \mbh\ and \ngc\ with stellar mass when the velocity dispersion is held fixed.  Thus, present observations lend evidence for feedback-regulated models in which the bulge binding energy is most important; they do not necessarily imply any `special' connection between globular clusters and \mbh.  This raises the question of why \ngc\ traces the formation of ellipticals and bulges sufficiently well to be correlated with binding energy.  
\end{abstract}


\section{Introduction} \label{s:intro}

There are now well-established correlations between the mass of supermassive black holes (SMBHs) and properties of their host galaxies, such as bulge luminosity, mass, light concentration, and velocity dispersion \citep{kr95, magorrian98, ferrarese00, gebhardt00, tremaine02, marconi03, gultekin09}.  This suggests that the physical mechanism driving growth of the SMBH also plays a key role in forming the bulge (for spiral galaxies) or galaxy (for ellipticals).  Analytical estimates \citep{sr98,burkert01,hh06}, as well as numerical simulations \citep{dsh05,sdh05,cox06_kinematics,robertson06_msigma,croton06,johansson09} with simple prescriptions for SMBH accretion have demonstrated the plausibility of this inference by matching the expected slopes of these correlations.  

Regardless of the detailed feedback prescription, these models predict that SMBHs grow until reaching some critical mass, where the energy and/or momentum released by feedback expels material from the nucleus. As such, they robustly predict that the ``true'' correlation should be between SMBH mass and a quantity such as the binding energy or potential well depth of material in the bulge. \citet{hopkinsfp_07a} show that the observed correlations with different variables, and importantly their scatter and systematic deviations from the relations, can be understood as the projections of a single fundamental dependence.  This relation is approximated closely by a multi-variable correlation, a black hole fundamental plane (BHFP). \citet{ar07} confirmed this in a sample of ellipticals and spiral bulges using dynamical models of bulge potentials, and \citet{fm09} did so with simple proxies such as $M_{BH} \propto E_b \sim M_* \sigma^2$.

Additional correlations have been found between SMBH mass and dark matter halo mass, as well as the number \ngc\ of globular clusters (GC) in the host galaxy \citep{sf09, bt10, harris10}.  In particular, \citet[][hereafter BT10]{bt10} argued that \ngc\ is a better predictor of \mbh\ than the velocity dispersion $\sigma$, citing a smaller intrinsic scatter and a residual correlation between \ngc\ and \mbh\ in elliptical galaxies even after accounting for the median $M_{BH}-\sigma$ correlation, suggesting a fundamental link between the accretion of gas by the SMBH and the formation of a galaxy's globular cluster system.  \citet[][hereafter HH10]{harris10} extended the sample by making reasonable estimates of \ngc\ from the literature in galaxies with \mbh\ measurements.  

In this letter, we illustrate that the above link can be understood as a consequence of the BHFP relation combined with a residual correlation between \ngc\ and the bulge's stellar mass $M_*$ at fixed $\sigma$.  Rather than suggesting a single ``best'' correlation between \mbh\ and a single galaxy parameter, the BHFP implies that the best predictor of SMBH mass is some combination thereof.  For example, \mbh\ has a positive correlation with the bulge's stellar mass even at fixed $\sigma$.  Although the number of globular clusters in a particular galaxy, like \mbh, is a complex function of the galaxy's formation history, there exists a similar positive residual correlation between \ngc\ and $M_*$, so that the resulting \ngc--\mbh\ residuals (fixing $\sigma$) will be positively correlated.  

In \S\ref{s:data} we describe a sample of 32 elliptical galaxies from \citet{peng08} with auxiliary data compiled in \citet[][and subsequent papers]{hopkins08_mr}.  In \S\ref{s:correlations} we fit separately the relations $M_*$--$\sigma$ and \ngc--$\sigma$ in these galaxies to establish the residual correlation between \ngc\ and $M_*$.  Then we combine this residual slope with knowledge of the \mbh--$M_*$ correlation at fixed $\sigma$ from the BHFP, and calculate the residual correlation and scatter expected between \ngc\ and \mbh.  We summarize and conclude in \S\ref{s:conclusions}.


\section{The Data} \label{s:data}

To determine the dependence of \ngc\ on $M_*$ at fixed $\sigma$, we cross-match objects compiled in \citet{hopkins08_mr} and subsequent works with the ACS Virgo Cluster Survey \citep[VCC,][]{cote04}, from which \citet{peng08} determined globular cluster counts (\ngc) and uncertainties.  Following BT10, we obtained \ngc\ for several additional galaxies from \citet{spitler08}.  

We obtained stellar masses and uncertainties from \citet{peng08} and \citet{hopkins09_cusp, hopkins09_core}, who compiled photometric data from several authors \citep[e.g.][and references therein]{bender88, rj04, lauer07a, kormendy09}.  We use velocity dispersions as compiled by \citet{hopkins09_cusp, hopkins09_core}. The latter quantity is the one best-determined for nearby massive galaxies, so we assume a log-uniform uncertainty in $\sigma$ of 0.02 dex, consistent with literature values.  This approach yields 33 galaxies for which we will determine the residual correlation between \ngc\ and $M_*$.  We discard the known recent merger remnant NGC1316 because its nuclear velocity dispersion is unrelaxed and the globular cluster system is actively evolving \citep{schweizer80}.  The galaxy properties used to analyze this 32-galaxy sample are provided in Table~\ref{table:1}.

In addition, we will utilize the \mbh, \ngc, and $\sigma$ data directly from Table 1 of BT10 (compiled mostly from \citet{gultekin09}), and the \mbh, \ngc\ data from Table 1 of HH10.  For the latter, we use the mean recorded $\sigma$ values from the Hyperleda database \citep{paturel03, mcelroy95}. These two sources yield 21 galaxies that serve as a combined comparison sample (Table~\ref{table:2}) for our derived \mbh--\ngc\ residual correlation.  We note that BT10 and HH10 use slightly different, but statistically consistent, values for \ngc\ where the samples overlap.  The differing values of \mbh\ may make a larger difference in cases where multiple measurements exist; here, we follow BT10 and give half weight to each in our fits.  Furthermore, different studies provide different values for the velocity dispersion of a given galaxy; for example, Hyperleda returns $\sigma$ values $\sim$10-20 km/s smaller than the ones from BT10, and the papers by Hopkins et al. provide values that differ by $\sim \pm$10-20 km/s.  We computed the observed residual \ngc--\mbh\ correlation using these alternate sources of $\sigma$, and find that the small changes this introduces leave our conclusions completely unchanged.

      \begin{figure*}
      \epsscale{1.0}
      \plotone{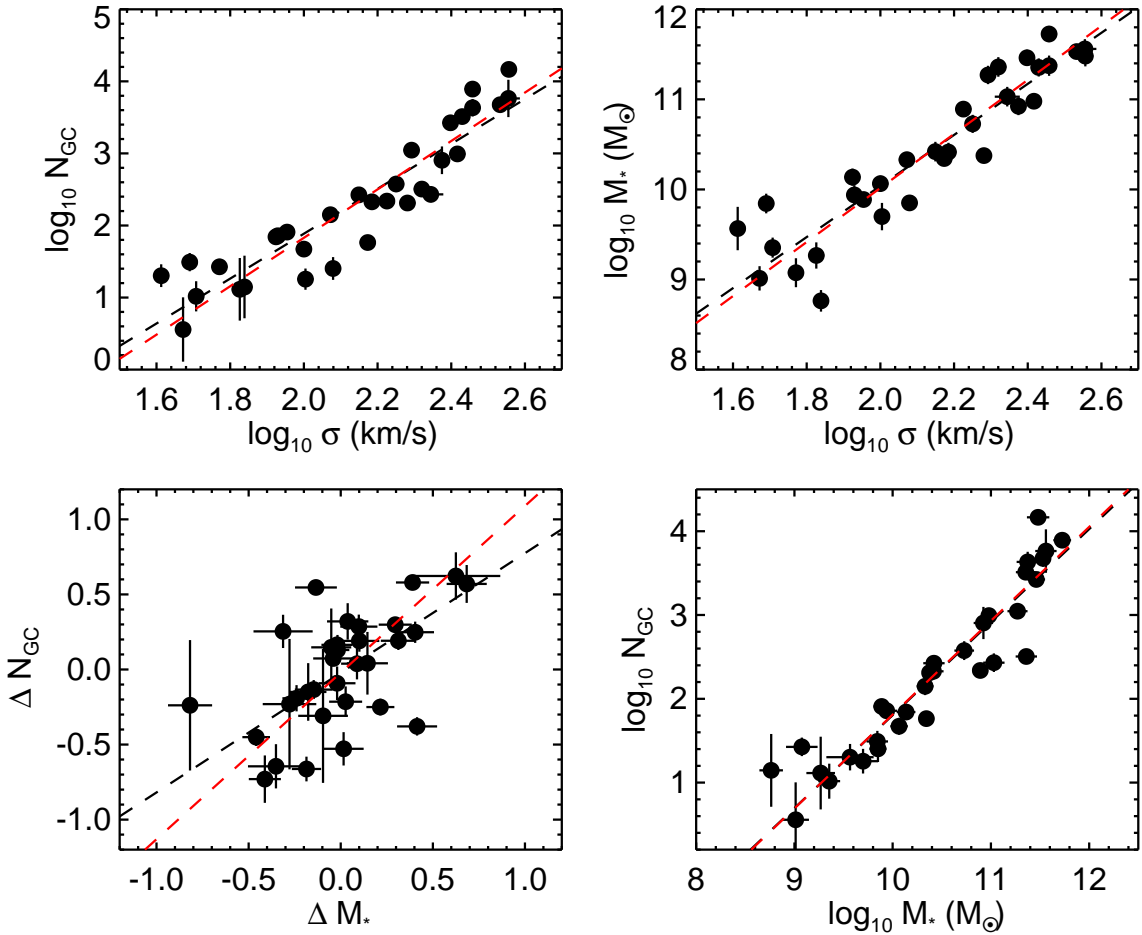}
      \caption{\emph{Top:} Observed correlations of \ngc\ and $M_*$ with velocity dispersion $\sigma$ for the 32-galaxy sample described in \S\ref{s:data}.  \emph{Lower Left:} Correlation between \ngc\ and $M_*$ at fixed $\sigma$: $\Delta $\ngc\ ($\Delta M_*$) is the difference between the observed logarithm of \ngc\ ($M_*$) and the expected value of the logarithm of \ngc\ ($M_*$) given the linear relation in the top left (top right) panel.  The lines are our regression curves fitted to the data.  The dashed black curves are fitted using a $\chi^2$ technique accounting for intrinsic scatter \citep{tremaine02}, while the dashed red curves use an alternative maximum-likelihood technique from \citet{ab96}.  There is a clear positive correlation between $\Delta$\ngc\ and $\Delta M_*$.  If the BHFP is the true underlying relation, then this observed residual correlation will lead to a correlation between \ngc\ and \mbh, even at fixed $\sigma$ and with no other physics linking \mbh\ to the globular cluster systems.  \emph{Lower Right: } The observed correlation between \ngc\ and $M_*$.   \label{fig:1} }
      \end{figure*}


\section{Correlations} \label{s:correlations}

Both the BHFP and \mbh--\ngc\ relation appear to perform better than the \mbh--$\sigma$ relation because they explain its residuals and hence have a smaller intrinsic dispersion.  Rigorously, this can be restated as follows: at fixed $\sigma$, the residuals in \mbh\ correlate tightly with the residuals in \ngc\ and in $M_{*}$ for the BHFP.  

The question then arises: does \ngc\ directly explain the residuals in \mbh, or could the latter be attributed to other variables already proposed?  Specifically, we examine whether the observed \ngc--\mbh\ relation is predicted as an indirect consequence of the BHFP relation.  This is motivated by the BHFP prediction that at fixed $\sigma$, good tracers of the bulge binding energy correlate tightly with the residuals in \mbh.  (Interestingly, a similar relation was shown for the binding energies of individual Milky Way globulars by \citet{mclaughlin00}.)  An approximation to the bulge binding energy in ellipticals is a quantity like $M_* \sigma^2$, so that at fixed $\sigma$, galaxies with larger $M_*$ will have a greater binding energy.  Thus, if \ngc\ adequately traces $M_*$ at fixed $\sigma$, as might be natural given the observed \ngc--galaxy correlations, then the tightness of the \ngc-\mbh\ relation is expected.  

In this work, we focus on a particular projection of the BHFP that uses the bulge binding energy as the driving parameter.  However, we note that the general BHFP, and also the corresponding salient relation for GCs \citep[e.g.][and subsequent works]{hvdb81, mclaughlin99} depends on galaxy formation history in a more complicated way \citep{hmt09}.  Thus while the bulge binding energy serves adequately for our purposes, a more detailed accounting of, for example, the total baryon mass may lead to an even tighter expected correspondence.  

As follows, we calculate the expected residuals in \ngc--\mbh\ assuming that this relation is a consequence of the BHFP and no other physics.  As an expression of the BHFP correlation, we use the relation between the mass of the SMBH and bulge binding energy from \citet{hopkinsfp_07a}, \[ \log M_{BH} = \eta + \beta \log (M_* \sigma^2),\] where $\eta = 8.23 \pm 0.06$, and $\beta = 0.71 \pm 0.06$.  We will denote this quantity as predicted from the other observed variables as $\log \langle$\mbh$|$BHFP$\rangle$.  Then we subtract from this the logarithm of \mbh\ as predicted solely from \mbh--$\sigma$, denoted by $\log \langle$\mbh$|\sigma \rangle$, to obtain the BHFP-predicted residual correlation between \mbh\ and $M_*$.  We will signify this difference in logarithmic quantities as $\Delta M_{BH}$ ($\Delta M_*$, $\Delta N_{GC}$): 
\begin{align*} \label{eq:1}
\Delta M_{BH} &= \log \langle M_{BH} | BHFP \rangle - \log \langle M_{BH} | \sigma \rangle = \\
&= \beta \log ( M_*\sigma^2 ) - \beta \log (\langle M_* | \sigma \rangle \sigma^2) \\
&= \beta ( \log M_* - \log \langle M_* | \sigma \rangle) = \beta \Delta M_*.  \tag{1}
\end{align*}  This is just the statement that at fixed $\sigma$, \mbh\ $\propto M_*^{\beta}$.  

If there exists a relation between $\Delta N_{GC}$ and $\Delta M_*$:
\begin{equation}\label{eq:2}
\Delta N_{GC} = \gamma \Delta M_*, \tag{2}
\end{equation}
this will therefore result in a correlation between $\Delta$\ngc\ and $\Delta$\mbh:
\begin{equation}\label{eq:3}
\Delta N_{GC} = \frac{\gamma}{\beta} \Delta M_{BH} = \alpha \Delta M_{BH}. \tag{3}
\end{equation}

\begin{figure*}
      \epsscale{0.75}
      \plotone{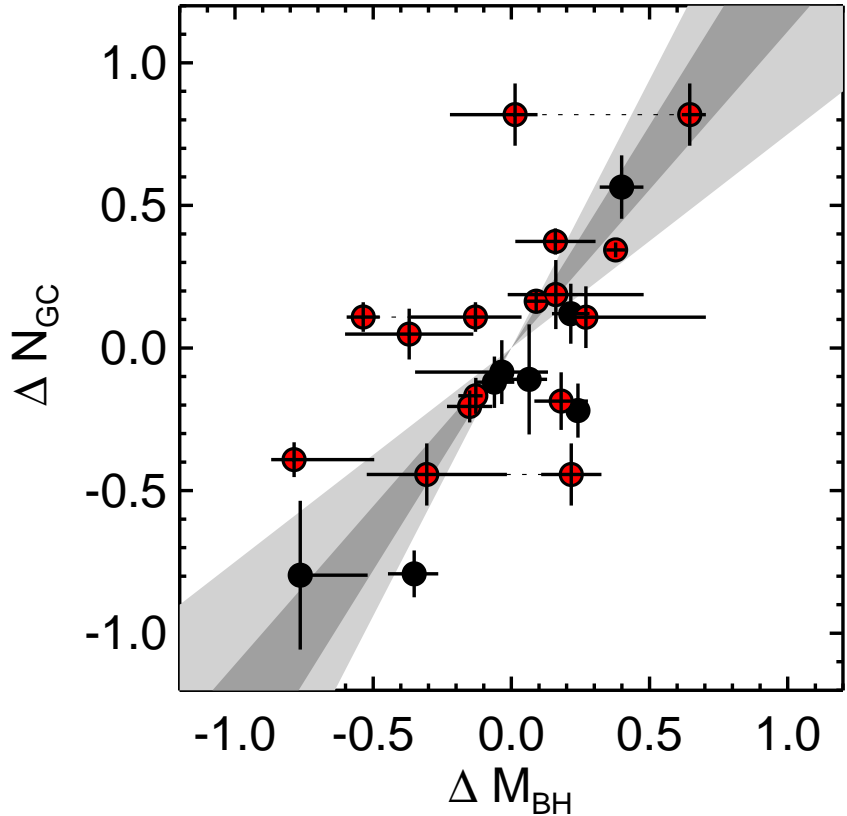}
      \caption{ Residual correlation between \ngc\ and \mbh\ at fixed velocity dispersion $\sigma$.  The points are data from BT10 and HH10; we follow BT10 where multiple \mbh\ measurements exist by assigning each measurement half weight in any fits.  Red points correspond to the BT10 sample; their positions here and in BT10's Figure 3 differ slightly because we take \ngc\ residuals with respect to $\sigma$ instead of bulge luminosity, but the residuals here have roughly the same slope and span a similar range of $\Delta$\ngc\ as compared with BT10.  The eight points added by the elliptical sample of HH10 reinforce this residual correlation and expand its range.  The dark gray shaded region is the {\em predicted} residual correlation (with associated 1-sigma uncertainty in lighter shade) assuming that \mbh\ is determined only by the BHFP, combined with the observed correlation between \ngc\ and $M_*$ at fixed $\sigma$ determined by fitting the data in Figure~\ref{fig:1}.  We see that the observed residual slope is in good agreement with this expected slope.  Thus the apparent additional predictive power of \ngc\ for \mbh\ can be entirely accounted for by the predicted correlation of \mbh\ and the observed correlation of \ngc\ with the bulge binging energy. \label{fig:2} }
      \end{figure*}

To test this, we first calculate $\gamma$ from the existing data to establish a correlation between \ngc\ and $M_*$ at fixed velocity dispersion $\sigma$ (Figure~\ref{fig:1}), and then combine it with the BHFP to create a prediction for the \ngc--\mbh\ residuals.  Then, in Figure~\ref{fig:2}, we compare this prediction to the observed residual correlation between \ngc\ and \mbh\ from BT10 and HH10.

Specifically, we begin by determining the observed best-fit linear correlations between $\log N_{GC}$ and $\log \sigma$, and $\log M_*$ and $\log \sigma$.  The resulting fits are shown in the top two panels of Figure~\ref{fig:1}.  We undertake all fits using two methods: the $\chi^2$-minimization methods of \citet{tremaine02} (hereafter, T02), and the bivariate correlated errors and intrinsic scatter (BCES) estimators of \citet{ab96}.  For the former, we account for intrinsic scatter in the Y axis by adding a uniform scatter in quadrature with the measurement errors such that the reduced-$\chi^2$ value of the fit is 1.  For the latter, we choose the regression line that bisects the BCES(Y$|$X) and BCES(X$|$Y) curves.  Altering these choices leads to small changes in the residual values, but does not change the residual slope in a statistically significant way.  One-sigma uncertainties in the fitted slopes are calculated using paired nonparametric bootstrap simulations \citep{br93}.

For each object, we then use the fitted relation to compute the expected value of $\log N_{GC}$ ($\log M_*$) given its observed value of $\sigma$, and subtract it from the observed value of $\log N_{GC}$ ($\log M_*$) to obtain $\Delta N_{GC}$ ($\Delta M_*$).  In the lower left panel of Figure~\ref{fig:1}, as expected we see a clear positive correlation between $\Delta$\ngc\ and $\Delta M_*$, indicating that at a fixed $\sigma$, elliptical galaxies with more globular clusters also have a larger total stellar mass.  We note that the values of $\Delta$\ngc\ and $\Delta$\mbh\ calculated using BCES or $\chi^2$-minimization on the direct correlations are the same to within 0.1 dex.  Subsequent estimates of the residual slope are unaffected by this choice.

However, the two fitting methods obtain somewhat different estimates for the slope of the resulting residual correlation, $\gamma$, which we will use to estimate the expected \ngc--\mbh\ residual slope $\alpha = \gamma/\beta$.  As plotted in Figure~\ref{fig:1}, we find
\begin{align*}
\gamma_{T02} &= 0.79 \pm 0.25 \\
\gamma_{BCES} &= 1.11 \pm 0.21,
\end{align*}
and correspondingly,
\begin{align*}
\alpha_{T02} &= 1.11 \pm 0.36 \\
\alpha_{BCES} &= 1.56 \pm 0.32.
\end{align*}

In Figure~\ref{fig:2}, we compare this predicted slope with the one observed in the comparison sample of BT10 and HH10.  Again, we compute the residuals against $\sigma$ in both \ngc\ and \mbh; note that for consistency, this is slightly different than the quantities plotted in BT10's Figure 3 where the \ngc\ residual was computed against the bulge luminosity, not velocity dispersion.  We plot a region in light gray to highlight the extremes of the predicted slopes, corresponding to the range bounded by the 1-sigma uncertainties in the slope given by our two regression methods.  In darker gray we simply plot the range bounded by our two slope estimates.  

As in Figure~\ref{fig:1}, we fit the data directly and find that the observed residual slope between \ngc\ and \mbh\ is
\begin{align*}
\hat{\alpha}_{T02} & = 0.78 \pm 0.24 \\
\hat{\alpha}_{BCES} &= 1.33 \pm 0.34,
\end{align*}
in good agreement with the BHFP predictions above.

The detection of this residual correlation by BT10 quantitatively demonstrates that \ngc\ is a better predictor of \mbh\ than is $\sigma$.  Such a comparison can be alternatively phrased as a reduction in the intrinsic scatter of the correlation.  In BT10, the intrinsic scatter of \mbh--$\sigma$ was found to be $\epsilon \sim 0.3$ dex, while the intrinsic scatter in \ngc--\mbh\ is $\epsilon \sim 0.2$ dex.  The magnitude of this dispersion can be predicted by combining the BHFP relation with the observed correlations with \ngc.  From \citet{hopkinsfp_07a}, we see that the \mbh--$E_b$ correlation has an intrinsic scatter $\sim 0.2$-$0.25$ dex, and from the present data, we find the scatter of $E_b$--\ngc\ is $0.22 \pm 0.04$ dex.  By propagating these as measurement uncertainties to the \mbh--\ngc\ relation, we predict that the measured \ngc--\mbh\ intrinsic scatter should be $\epsilon = 0.23 \pm 0.03$ dex, consistent with the measurement by BT10.  It is also consistent with the combined dataset of BT10 and HH10, for which we find $\epsilon = 0.21 \pm 0.04$ dex.


\section{Conclusions} \label{s:conclusions}

We have shown that the number of globular clusters in elliptical galaxies exhibits a residual dependence on $M_*$ at fixed $\sigma$, implying that the bulge binding energy ($\sim M_* \sigma^2$) is a better indicator of \ngc\ than $\sigma$ or $M_*$ alone.  The same was shown to be true for \mbh\ by \citet{hopkinsfp_07a}, as these parameters constitute a formulation of the BHFP.  Thus the apparent power of \mbh--\ngc\ versus \mbh--$\sigma$ owes to the fact that \ngc\ and \mbh\ are both tracers of the same fundamental property such as the bulge binding energy.  

This resolves several puzzling aspects of the previous interpretation of the data.  As BT10 themselves point out, there cannot be a direct causal correlation between \ngc\ and \mbh, since most of the GC mass is at very large radii and has never had any interaction with the galaxy nucleus.  Moreover, while most GCs likely formed at very high redshift, the final mass of the SMBH is sensitive to its growth via gas accretion at $z\lesssim2$ \citep[e.g.][]{hh06,hrh07,hopkins08_ell}.  However, this naturally predicts that \ngc\ should serve reasonably well as a mass tracer, so that the dependence of \mbh\ on $M_*$ and formation time leads to a surprisingly tight but expected \ngc--\mbh\ correlation.  The same arguments explain the result in \citet{hmt09}, who show that the observed \mbh\ is sensitive to the entire galaxy baryonic mass -- i.e. perhaps the mass traced by \ngc\ is the same as the mass that actually sets the escape velocity and potential well depth at $R=0$, rather than just the stellar mass enclosed in a small radius around the BH, which can vary widely in systems of similar \mbh.  Such a relation between \ngc\ and global galaxy mass or luminosity has been demonstrated \citep[e.g.][]{hvdb81, mclaughlin99}, and this trait supports the idea that \ngc\ and \mbh\ are connected indirectly by a more fundamental galaxy property.  

This also naturally explains why HH10 find that the relation breaks down for S0 galaxies.  These galaxies are structurally different than ellipticals and may have different formation histories \citep{larson80}, so \ngc\ and the total stellar mass may not faithfully trace the bulge binding energy.  Since S0's are not particularly discrepant in \mbh--$\sigma$ \citep[e.g.][and previous works]{gultekin09}, this suggests that the \ngc--bulge relation is the connection that weakens for these systems.  HH10 also find no statistically significant correlation in spirals: although three out of the four spirals from HH10 lie on the \ngc--\mbh\ relation, there simply isn't yet enough data to know for sure if this relation persists for spiral bulges.  However, the underlying BHFP relation ties \mbh\ to the binding energy and explains its residual correlations with bulge parameters, even for these disky galaxies where \ngc\ possibly deviates.  This alone suggests that the BHFP, not \mbh--\ngc\ or \mbh--$\sigma$, is the `more fundamental' correlation.  

\citet{hopkinsfp_07b} showed that the existence of a black hole fundamental plane is a robust prediction of numerical simulations of gas-rich mergers that include the effects of gas dissipation, cooling, star formation, and black hole accretion and feedback.  The present work shows that this local and widely expected correlation between supermassive black hole mass and bulge binding energy in feedback-regulated scenarios, combined with a similar correlation for \ngc, can account for the observed \ngc--\mbh\ relation and its scatter.  The interesting question raised by such a correlation is {\em not} why \ngc\ correlates tightly with \mbh, since this is indirect, but why \ngc\ correlates tightly with galaxy binding energy/potential well depth.  Some such correlation is expected and observed \citep{mclaughlin99, blakeslee99, peng08}: an example is that systems at fixed velocity dispersion with higher stellar mass have accreted or formed more stars, likely including globular clusters.  But that the \ngc--bulge relation should be so tight, and include both metal-rich and metal-poor populations, may support the inferences by BT10 and HH10 (and references therein) that the formation of globular cluster systems and growth of supermassive black holes in elliptical galaxies are driven by a common galaxy property.  

\acknowledgements

We thank the anonymous referees for numerous helpful comments.  




\clearpage

\begin{deluxetable}{cccc}

         
         
         
\tablecaption{Galaxy properties for Figure~\ref{fig:1}}
         
\tablenum{1}
         
\tablecolumns{4}
\tablehead{\colhead{Galaxy} & \colhead{\ngc} & \colhead{$M_*$} & \colhead{$\sigma$} \\ 
\colhead{} & \colhead{} & \colhead{($10^9 M_{\odot}$)} & \colhead{(km/s)} } 
         
\startdata
NGC0821 & 320 $\pm$ 45\tablenotemark{a} & 229 $\pm$ 57 & 209 $\pm$ 10 \\
NGC1399 & 5800 $\pm$ 700\tablenotemark{a} & 363 $\pm$ 91 & 359 $\pm$ 18 \\
NGC3377 & 266 $\pm$ 66\tablenotemark{b} & 26.3 $\pm$ 6.6 & 141 $\pm$ 7 \\
NGC3379 & 270 $\pm$ 68\tablenotemark{a} & 107 $\pm$ 27 & 221 $\pm$ 11 \\
NGC4318 & 18 $\pm$ 6.1 & 5.0 $\pm$ 1.7 & 101 $\pm$ 5 \\
NGC4365 & 3246 $\pm$ 598 & 226 $\pm$ 52 & 269 $\pm$ 13 \\
NGC4374 & 4301 $\pm$ 1201 & 236 $\pm$ 61 & 287 $\pm$ 14 \\
NGC4382 & 1110 $\pm$ 181 & 186 $\pm$ 44 & 196 $\pm$ 10 \\
NGC4387 & 69.5 $\pm$ 9.8 & 13.7 $\pm$ 3 & 84 $\pm$ 4.2 \\
NGC4406 & 2660 $\pm$ 129 & 289 $\pm$ 60 & 250 $\pm$ 12 \\
NGC4434 & 141 $\pm$ 34 & 21.4 $\pm$ 4 & 118 $\pm$ 6 \\
NGC4458 & 72 $\pm$ 12 & 8.7 $\pm$ 2 & 85 $\pm$ 4.3 \\
NGC4459 & 218 $\pm$ 28 & 77.9 $\pm$ 14 & 168 $\pm$ 8 \\
NGC4464 & 25.3 $\pm$ 9.2 & 7.1 $\pm$ 1.4 & 120 $\pm$ 6 \\
NGC4467 & -6 $\pm$ 13 & 1.8 $\pm$ 0.6 & 67 $\pm$ 3.4 \\
NGC4472 & 7813 $\pm$ 830 & 531 $\pm$ 110 & 287 $\pm$ 14 \\
NGC4473 & 76 $\pm$ 97 & 53.5 $\pm$ 12 & 178 $\pm$ 9 \\
NGC4476 & 20.1 $\pm$ 7.3 & 3.7 $\pm$ 2 & 41 $\pm$ 2.1 \\
NGC4478 & 58 $\pm$ 11 & 22 $\pm$ 4 & 149 $\pm$ 7 \\
NGC4486 & 14660 $\pm$ 891 & 302 $\pm$ 79 & 360 $\pm$ 18 \\
NGC4489 & 31 $\pm$ 9 & 6.98 $\pm$ 1.7 & 49 $\pm$ 2.4 \\
NGC4515 & 81 $\pm$ 10 & 7.7 $\pm$ 1.5 & 90 $\pm$ 4.5 \\
NGC4551 & 47 $\pm$ 11 & 11.6 $\pm$ 2.4 & 100 $\pm$ 5 \\
NGC4552 & 984 $\pm$ 198 & 95 $\pm$ 16.9 & 261 $\pm$ 13 \\
NGC4564 & 213 $\pm$ 31 & 26 $\pm$ 6 & 153 $\pm$ 8 \\
NGC4621 & 803 $\pm$ 355 & 83.9 $\pm$ 19 & 237 $\pm$ 12 \\
NGC4649 & 4745 $\pm$ 1099 & 339 $\pm$ 50 & 341 $\pm$ 17 \\
NGC4660 & 205 $\pm$ 28 & 23.8 $\pm$ 4 & 191 $\pm$ 9 \\
VCC1199 & -9 $\pm$ 14 & 0.58 $\pm$ 0.16 & 69 $\pm$ 3.5 \\
VCC1440 & 26.7 $\pm$ 6.8 & 1.2 $\pm$ 0.44 & 59 $\pm$ 3 \\
VCC1627 & 3.6 $\pm$ 3.7 & 1.0 $\pm$ 0.32 & 47 $\pm$ 2.4 \\
VCC1871 & 10.4 $\pm$ 5 & 2.3 $\pm$ 0.58 & 51 $\pm$ 2.6 \\
\enddata 
         
\tablenotetext{a}{\citet{spitler08}}
\tablenotetext{b}{\citet{kw01}}
         
         
\tablecomments{Properties of elliptical galaxies used to determine the correlation of \ngc\ with $M_*$ at fixed velocity dispersion $\sigma$ (Figure~\ref{fig:1}).  Values of \ngc\ and $M_*$ are compiled from \citet{peng08} unless otherwise noted.  Values of $\sigma$ are compiled from \citet{hopkins09_cusp,hopkins09_core} and assumed to have a log-uniform uncertainty of 0.02 dex.  \label{table:1}}

         
\end{deluxetable}

\clearpage
\begin{deluxetable}{cccccccc}
         



\tablecaption{Galaxy properties for Figure~\ref{fig:2}}
         
\tablenum{2}

\tablecolumns{8}
\tablehead{\colhead{Galaxy} & \colhead{\mbh} & \colhead{+1-sigma} & \colhead{-1-sigma} & \colhead{\ngc} & \colhead{$\sigma$} & \colhead{$\Delta$\ngc}  & \colhead{$\Delta$\mbh} \\ 
\colhead{} & \colhead{($M_{\odot}$)} & \colhead{($M_{\odot}$)} & \colhead{($M_{\odot}$)} & \colhead{} & \colhead{(km/s)} & \colhead{} & \colhead{}} 
         
\startdata
\sidehead{\citet{bt10}}
NGC0821 & $4.2\times 10^7$ & $2.8\times 10^7$ & $8\times 10^6$ & 320 $\pm$ 45 & 209 $\pm$ 10 & -0.39 & -0.79 \\
NGC1316 & $1.5\times 10^8$ & $8\times 10^7$ & $8\times 10^7$ & 1173 $\pm$ 240 & 226 $\pm$  9 & 0.05 & -0.37 \\
NGC1399 & $1.3\times 10^9$ & $5\times 10^8$ & $7\times 10^8$ & 5800 $\pm$ 700 & 337 $\pm$ 16 & 0.11 & -0.13 \\
\phm{7} & $5.1\times 10^8$ & $7\times 10^7$ & $7\times 10^7$ & \phn & \phn & 0.11 & -0.54 \\
NGC3377 & $1.1\times 10^8$ & $1.1\times 10^8$ & $1\times 10^7$ & 266 $\pm$ 66 & 145 $\pm$  7 & 0.11 & 0.27 \\
NGC3379 & $1.2\times 10^8$ & $8\times 10^7$ & $6\times 10^7$ & 270 $\pm$ 68 & 206 $\pm$ 10 & -0.44 & -0.31 \\
\phm{7} & $4\times 10^8$ & $1\times 10^8$ & $1\times 10^8$ & \phn & \phn & -0.44 & 0.22 \\
NGC4374 & $1.5\times 10^9$ & $1.1\times 10^9$ & $6\times 10^8$ & 4301 $\pm$ 1201 & 296 $\pm$ 14 & 0.19 & 0.16 \\
NGC4459 & $7.4\times 10^7$ & $1.4\times 10^7$ & $1.4\times 10^7$ & 218 $\pm$ 28 & 167 $\pm$  8 & -0.20 & -0.15 \\
NGC4472 & $1.8\times 10^9$ & $6\times 10^8$ & $6\times 10^8$ & 7813 $\pm$ 830 & 310 $\pm$ 10 & 0.37 & 0.16 \\
NGC4486 & $6.4\times 10^9$ & $5\times 10^8$ & $5\times 10^8$ & 14660 $\pm$ 891 & 375 $\pm$ 18 & 0.34 & 0.38 \\
NGC4564 & $6.9\times 10^7$ & $4\times 10^6$ & $1\times 10^7$ & 213 $\pm$ 31 & 162 $\pm$  8 & -0.17 & -0.13 \\
NGC4594 & $5.5\times 10^8$ & $5\times 10^7$ & $5\times 10^7$ & 1900 $\pm$ 189 & 240 $\pm$ 12 & 0.16 & 0.09 \\
NGC4649 & $4.5\times 10^9$ & $1\times 10^9$ & $1\times 10^9$ & 4745 $\pm$ 1099 & 385 $\pm$ 19 & -0.19 & 0.18 \\
NGC5128 & $3\times 10^8$ & $4\times 10^7$ & $2\times 10^7$ & 1550 $\pm$ 390 & 150 $\pm$  7 & 0.82 & 0.65 \\
\phm{7} & $7\times 10^7$ & $1.3\times 10^7$ & $3.8\times 10^7$ & \phn & \phn & 0.82 & 0.01 \\
\sidehead{\citet{harris10}}
NGC2778 & $1.6\times 10^7$ & $9\times 10^6$ & $2\times 10^5$ & 50 $\pm$ 30 & 162 $\pm$  8 & -0.80 & -0.76 \\
NGC4261 & $5.5\times 10^8$ & $1.1\times 10^8$ & $1.2\times 10^8$ & 530 $\pm$ 100 & 309 $\pm$ 14 & -0.79 & -0.35 \\
NGC4473 & $1.3\times 10^8$ & $5\times 10^7$ & $9.4\times 10^7$ & 376 $\pm$ 97 & 180 $\pm$  8 & -0.08 & -0.03 \\
NGC4552 & $4.8\times 10^8$ & $8\times 10^7$ & $8\times 10^7$ & 1200 $\pm$ 250 & 253 $\pm$ 12 & -0.12 & -0.06 \\
NGC4621 & $4\times 10^8$ & $6\times 10^7$ & $6\times 10^7$ & 800 $\pm$ 355 & 225 $\pm$ 11 & -0.11 & -0.06 \\
NGC4697 & $2\times 10^8$ & $2\times 10^7$ & $2\times 10^7$ & 229 $\pm$ 50 & 171 $\pm$  8 & -0.22 & 0.24 \\
NGC5813 & $7\times 10^8$ & $1.1\times 10^8$ & $1.1\times 10^8$ & 1650 $\pm$ 400 & 237 $\pm$ 11 & 0.12 & 0.22 \\
NGC5846 & $1.1\times 10^9$ & $2\times 10^8$ & $2\times 10^8$ & 4700 $\pm$ 1200 & 239 $\pm$ 11 & 0.56 & 0.40 \\
\enddata 
         
\tablecomments{Properties of elliptical galaxies used to calculate the correlation of \ngc\ with \mbh\ at fixed velocity dispersion $\sigma$ (Figure~\ref{fig:2}), following BT10.  Data for \ngc\ and \mbh\ are compiled from BT10 and HH10.  Values of $\sigma$ are as presented in BT10 for those galaxies, and values of $\sigma$ for the HH10 ellipticals are taken as the mean value recorded in the HyperLeda database \citep{paturel03, mcelroy95}. \label{table:2}}
         



\end{deluxetable}

\clearpage
\end{document}